\documentclass[aps, prd,twocolumn, showpacs, showkeys,nofootinbib,floatfix]{revtex4-1}

\usepackage{amssymb}
\usepackage{amsmath}
\usepackage{graphicx}
\usepackage{hyperref}

\begin{document}

\title{Equation of State of Dark Matter after Planck Data}
\author{Lixin Xu}
\email{lxxu@dlut.edu.cn}
\author{Yadong Chang}

\affiliation{Institute of Theoretical Physics, School of Physics \&
Optoelectronic Technology, Dalian University of Technology, Dalian,
116024, P. R. China}

\affiliation{College of Advanced Science \& Technology, Dalian
University of Technology, Dalian, 116024, P. R. China}

\date{\today}

\begin{abstract}
In this paper, we loosen the zero equation of state of dark matter to a constant $w_{dm}$. By using the currently available cosmic observations which include the type-Ia supernovae, the baryon acoustic oscillation, the WiggleZ measurements of matter power spectrum and the cosmic microwave background radiation from the firs release of {\it Planck} dat through the Markov chain Monte Carlo method, we found the equation of state of dark matter in $3\sigma$ regions: $w_{dm}=0.000707_{-0.000747-0.00149-0.00196}^{+0.000746+0.00146+0.00193}$. The difference of the minimum $\chi^2$ between the $\Lambda$CDM and $\Lambda$wDM models is $\Delta\chi^2_{min}=0.446$ for one extra model parameter $w_{dm}$. Although the currently available cosmic observations favor the $\Lambda$wDM mildly, no significant deviation from the $\Lambda$CDM model is found in $1\sigma$ regions.  	 
\end{abstract}



\maketitle

\section{Introduction}

The firs release of {\it Planck} data improves the quality of cosmological data extraordinary \cite{ref:Planck}. It allows us to give a tighter constraint to the cosmological parameter space. And the cold dark matter plus a cosmological constant $\Lambda$, the so-called $\Lambda$CDM model, can almost agree with the most recent cosmic observations which include the type-Ia supernovae (SN), the baryon acoustic oscillation (BAO) and the cosmic microwave background radiation (CMB) successfully at large scales. However, it has several potential problems on smaller scales \cite{Klypin1999,Moore1999,Zavala2009,Tikhonov2009,deBlok2001}. How to explain the discrepancies on large and small scales is currently still under debate \cite{Weinberg2013}. The warm dark matter has been proclaimed as a potential solution to the small scale difficulties of cold dark matter \cite{Zavala2009,Boylan-Kolchin2012,Lovell2012,Anderhalden2013,Menci2013,Papastergis2011}. It leaves some space to an alternative to the cold dark matter model. The focus point is whether it is cold or warm. Actually, the hot dark matter was ruled out due to the difficulty in forming the observed large scale structure. To characterize the properties of warm dark matter, the equation of state $w_{dm}$ is an important indicator in a fluid perspective. And the value of $w_m$ should be determined by the cosmic observations. A significant nonzero value of $w_{dm}$ indicates the dark matter is warm rather than cold. In the literature, the equation of state of WDM was constrained by many groups under the assumption of a constant $w_{dm}$, a time variable $w_{dm}$ with a cosmological constant or $w=constant$ dark energy. For an example, please see \cite{ref:Muller2005,ref:Calabrese2009,ref:Wei2013} and references therein. See also Ref. \cite{ref:Schneider2013}, in which the difficulties of the small scale behavior of warm dark matter were pointed out. If the dark matter is really warm the model parameters would be different even it were to have the same background evolution history as that of the $\Lambda$CDM model. A different large scale structure would form due to the different perturbation evolutions. Therefore, the large scale structure information will be important to break degeneracies between model parameters and to determine the properties of dark matter which is the main part of the large scale structure of our Universe. We would also like to mention the particle side of dark matter; please see Ref. \cite{ref:WDMParticleReview} for brief reviews.

Thanks to the measurements of WiggleZ Dark Energy Survey, a total $238,000$ galaxies in the redshift range $z<1$ were measured. These galaxies were split into four redshift bins with ranges $0.1<z<0.3$, $0.3<z<0.5$, $0.5<z<0.7$ and $0.7<z<0.9$. The corresponding power spectrum in the four redshift bins was measured; for details, please see \cite{ref:WiggleZ2012}. We estimate the nonlinear growth from a given linear growth theory power spectrum based on the principles of the halo model; actually the HALOFIT formula will be used in this paper \cite{ref:HALOFIT}. At the scale of halo, the growth of halos depends on the local physics, and not on the details of precollapse matter and the large scale distribution of matter. Thus, in the nonlinear regime the growth depends only on the nonlinear scale, the slope and curvature of the power spectrum \cite{ref:Bird2011}. In this paper, we loosen the constraint to a zero equation of state and investigate the simplest model for dark matter, i.e. the one with a constant $w_{dm}$. Therefore, the main information is stored in the matter power spectrum. Also we assume the HALOFIT formula is still suitable for this case, though the formula would be modified due to the free-streaming of warm dark matter \cite{ref:Smith2011}. By a combination of CMB, SDSS BAO, SN and WiggleZ, the EoS of dark matter will be tested. As results, we did not find significant deviation from $w_{dm}=0$ in $1\sigma$ regions.         

Actually, this simplest form dark matter was already constrained by using $580$ SN, CMB shift parameter $R$ and BAO distance parameter $A$ \cite{ref:Wei2013} (see also in Refs. \cite{ref:Muller2005,ref:Calabrese2009}), in which a time variable equation os state of dark matter was also considered. The authors of Ref. \cite{ref:Calabrese2009} discussed the degeneracies between model parameters extensively and found the equation os state of dark matter in the range of $1-2\sigma$ regions $w_{dm}=0.0007^{+0.0021+0.0041}_{-0.0021-0.0042}$ by using the data combination of WMAP5+SDSS+SNLS. It implies the cold dark matter $w_{dm}=0$ is compatible in the $1\sigma$ region. As a revisit to the work of Ref. \cite{ref:Calabrese2009} and a comparison to the work of Ref. \cite{ref:Wei2013}, here we will use the full information of CMB, which includes the recently released {\it Planck} data sets, which include the high-l TT likelihood ({\it CAMSpec}) up to a maximum multipole number of $l_{max}=2500$ from $l=50$; the low-l TT likelihood ({\it lowl}) up to $l=49$; and the low-l TE, EE, BB likelihoods up to $l=32$ from WMAP9; the data sets are available on line \cite{ref:Planckdata}. For the SN data points as "standard candles", the luminosity distances will be employed. In this paper, we keep to use the SNLS3 which consists of $472$ SN calibrated by SiFTO and SALT2; for details, please see Ref. \cite{ref:SNLS3}. Although the photometric calibration of the SNLS and the SDSS Supernova Surveys were improved \cite{ref:SNLS3recal}, they are still unavailable publicly. For the BAO data points as "standard ruler", we use the measured ratio of $D_V/r_s$, where $r_s$ is the comoving sound horizon scale at the recombination epoch, and $D_V$ is the "volume distance" which is defined as
\begin{equation}
D_V(z)=[(1+z)^2D^2_A(z)cz/H(z)]^{1/3},
\end{equation}
where $D_A$ is the angular diameter distance. Here the BAO measurements from WiggleZ are not included, as they come from the same galaxy sample as the $P(k)$ measurement. Because the full information of CMB and WiggleZ power spectrum is employed, a tiger constraint is expected. Our results will show that it is indeed the case. 

The plan of this paper is as follows: in Sec. \ref{sec:BPEs}, we present the main background evolution and perturbation equations for dark matter with an arbitrary equation os state. In Sec. \ref{sec:results}, the constrained results are presented via the Markov chain Monte Carlo (MCMC) method. Section \ref{sec:con} is the conclusion. 

\section{Background and Perturbation equations} \label{sec:BPEs}

The equation os state of dark matter is given as
\begin{equation}
w_{dm}=\frac{p_{dm}}{\rho_{dm}}.
\end{equation}
The Friedmann equation for a spatially flat FRW universe reads
\begin{equation}
H^2=H^2_0\left[\Omega_{r}a^{-4}+\Omega_{b}a^{-3}+\Omega_{dm}a^{-3(1+w_{dm})}+\Omega_{\Lambda}\right],\label{eq:EHFE}
\end{equation}
where $\Omega_{i}=\rho_i/3 M^2_{pl}H^2$ are the present dimensionless energy densities for the radiation, the baryon, the dark matter and the cosmological constant respectively, where $\Omega_{dm}+\Omega_{r}+\Omega_{b}+\Omega_{\Lambda}=1$ is respected for a spatially flat universe.

In this paper, the dark matter is taken as a perfect fluid with a constant equation os state, and then in the synchronous gauge the perturbation equations of density contrast and velocity divergence for the dark matter are written as
\begin{eqnarray}
\dot{\delta}_{dm}&=&-(1+w_{dm})(\theta_{dm}+\frac{\dot{h}}{2})-3\mathcal{H}(\frac{\delta p_{dm}}{\delta \rho_{dm}}-w_{dm})\delta_{dm},\label{eq:continue}\\
\dot{\theta}_{dm}&=&-\mathcal{H}(1-3c^2_{s,ad})+\frac{\delta p_{dm}/\delta \rho_{dm}}{1+w_{dm}}k^{2}\delta_{dm}-k^{2}\sigma_{dm}\label{eq:euler}
\end{eqnarray}
following the notations of Ma and Bertschinger \cite{ref:MB}, in which the definition of the adiabatic sound speed
\begin{equation}
c^2_{s,ad}=\frac{\dot{p}_{dm}}{\dot{\rho}_{dm}}=w_{dm}-\frac{\dot{w}_{dm}}{3\mathcal{H}(1+w_{dm})}
\end{equation}
 is used. Here the equations are also ready for a general form of dark matter. When the equation os state of a pure barotropic fluid is negative, the imaginary adiabatic sound speed can cause instability of the perturbations. To overcome this problem, one can introduce an entropy perturbation and assume a positive or null effective speed of sound. Following the work of Ref. \cite{ref:Hu98}, the nonadiabatic stress or entropy perturbation can be separated out
 \begin{equation}
 p_{dm}\Gamma_{dm}=\delta p_{dm}-c^2_{s,ad}\delta \rho_{dm}, \label{eq:entropyper}
\end{equation} 
which is gauge independent. In the rest frame of dark matter, the entropy perturbation is specified as
 \begin{equation}
 w_{dm}\Gamma_{dm}=(c^2_{s,eff}-c^2_{s,ad})\delta^{rest}_{dm},\label{eq:restframe}
 \end{equation}
where $c^2_{s,eff}$ is the effective speed of sound. Transforming into an arbitrary gauge 
 \begin{equation}
 \delta^{rest}_{dm}=\delta_{dm}+3\mathcal{H}(1+w_{dm})\frac{\theta_{dm}}{k^2}\label{eq:gaugetrans}
 \end{equation}  
 gives a gauge-invariant form for the entropy perturbations. By using Eqs (\ref{eq:entropyper},) (\ref{eq:restframe}) and (\ref{eq:gaugetrans}), one can recast Eqs. (\ref{eq:continue}), and (\ref{eq:euler}) into 
 \begin{widetext}
 \begin{eqnarray}
 \dot{\delta}_{dm}&=&-(1+w_{dm})(\theta_{dm}+\frac{\dot{h}}{2})+\frac{\dot{w}_{dm}}{1+w_{dm}}\delta_{dm}-3\mathcal{H}(c^2_{s,eff}-c^2_{s,ad})\left[\delta_{dm}+3\mathcal{H}(1+w_{dm})\frac{\theta_{dm}}{k^2}\right],\label{eq:wdmdelta}\\
\dot{\theta}_{dm}&=&-\mathcal{H}(1-3c^2_{s,eff})\theta_{dm}+\frac{c^2_{s,eff}}{1+w_{dm}}k^2\delta_{dm}-k^2\sigma_{dm}\label{eq:wdmv}.
 \end{eqnarray}
\end{widetext}
For the dark matter, we assume the shear perturbation $\sigma_{dm}=0$ and the adiabatic initial conditions. Actually, the effective speed of sound $c^2_{s,eff}$ is another freedom to describe the microscale property of dark matter in addition to the equation os state. And, we should take it as another free model parameter. The sound speed determines the sound horizon of the fluid via the equation $l_s=c_{s,eff}/H$. The fluid can be smooth or cluster below or above the sound horizon $l_s$ respectively. If the sound speed is smaller, the perturbation of the fluid can be detectable on large scale. In turn the clustering fluid can influence the growth of density perturbations of matter, large scale structure and evolving gravitational potential which generates the integrated Sachs-Wolfe effects. The dark matter is responsible to formation of the large scale structure of our Universe, so we assume the effective speed of sound $c^2_{s,eff}=0$ in this work. 

\section{Constrained Results} \label{sec:results}

To obtain the equation os state of dark matter from the currently available cosmic observations, we use the MCMC method which is efficient in the case of more parameters. We modified the publicly available {\bf cosmoMC} package \cite{ref:MCMC} to include the perturbation evolutions of dark matter with a general form of equation os state according to the Eqs. (\ref{eq:wdmdelta}) and (\ref{eq:wdmv}). We adopted the seven-dimensional parameter space
\begin{equation}
P\equiv\{\omega_{b},\omega_c, \Theta_{S},\tau, w_{dm}, n_{s},\log[10^{10}A_{s}]\}
\end{equation}
the priors for the model parameters are summarized in Table \ref{tab:results}. Furthermore, the hard-coded prior on the comic age $10\text{Gyr}<t_{0}<\text{20Gyr}$ is also imposed. Also, the new Hubble constant $H_{0}=72.0\pm3.0\text{kms}^{-1}\text{Mpc}^{-1}$ \cite{ref:hubble} is adopted. The pivot scale of the initial scalar power spectrum $k_{s0}=0.05\text{Mpc}^{-1}$ is used in this paper.

\begin{center}
\begin{figure}[htb]
\includegraphics[width=8.5cm]{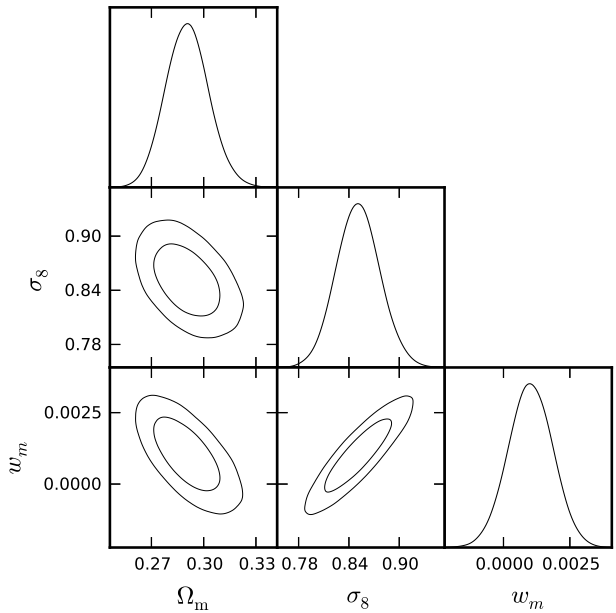}
\caption{The one-dimensional marginalized distribution on individual parameters and two-dimensional contours  with $68\%$ C.L., $95\%$ C.L. for the $\Lambda$wDM model by using CMB+BAO+SN data points.}\label{fig:wnowigglecontoure}
\end{figure}
\end{center}

\begin{center}
\begin{figure}[htb]
\includegraphics[width=8.5cm]{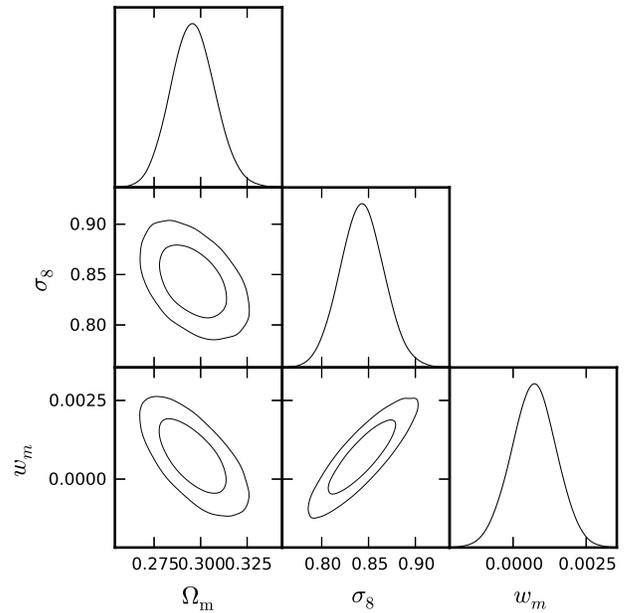}
\caption{The same as Fig. \ref{fig:wnowigglecontoure}, but with the WiggleZ measurements of the power spectrum. }\label{fig:wcontoure}
\end{figure}
\end{center}

\begin{center}
\begin{figure}[htb]
\includegraphics[width=8.5cm]{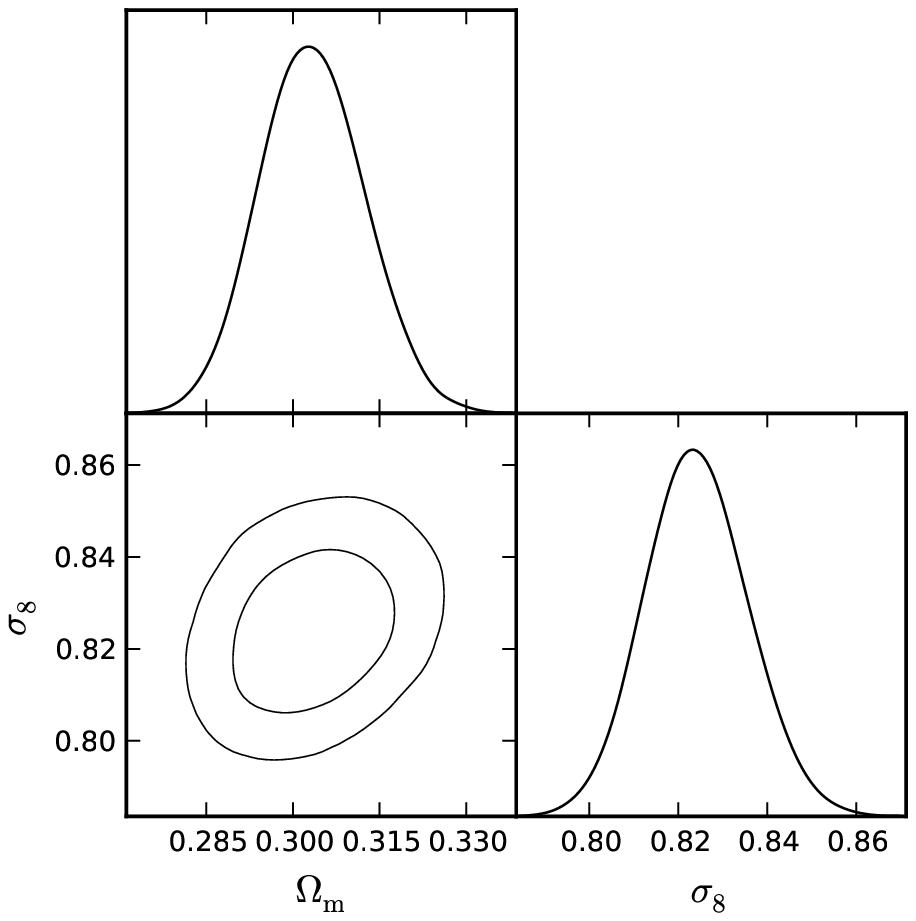}
\caption{The same as Fig. \ref{fig:wnowigglecontoure}, but with the WiggleZ measurements of the power spectrum for the $\Lambda$CDM model.}\label{fig:lcdmcontoure}
\end{figure}
\end{center}

After running eight chains for every cosmological model with different values of $w_{dm}$ on the {\it Computing Cluster for Cosmos}, we show the obtained results in Table \ref{tab:results}. This result is compatible with the previous results as shown in Refs. \cite{ref:Muller2005,ref:Calabrese2009,ref:Wei2013}; however a tighter constraint was obtained due to the high quality of current cosmic observational data points and the inclusion of WiggleZ measurements of the power spectrum. The $1-D$ and $2-D$ contours for $\Omega_m$, $\sigma_8$ and $w_{dm}$ were plotted in Figs. \ref{fig:wnowigglecontoure} and \ref{fig:wcontoure}, where the degeneracy between model parameter $\Omega_m$ and $w_{dm}$ can be understood easily: the amount of $\Omega_m$ at the decoupling epoch is fixed by the CMB power spectra; to maintain the same early time values of $\Omega_m$, lower present values of $\Omega_m$ need larger values of $w_{dm}$. This implies that the determination of $\Omega_m$ is vital to pin down the equation os state of dark matter. The inclusion of WiggleZ measurements of power spectrum decreases the values of $w_{dm}$ and $\sigma_8$ and shrinks the bounds mildly. For comparison, by using the same data sets combination, the model parameters of the $\Lambda$CDM model were also obtained as shown in Table \ref{tab:results} and Fig. \ref{fig:lcdmcontoure}. The difference of the minimum $\chi^2$ between the $\Lambda$CDM and $\Lambda$wDM models is $\Delta\chi^2_{min}=0.446$ for one extra model parameter $w_{dm}$. Although the currently available cosmic observations favor the $\Lambda$wDM mildly, no significant deviation from $\Lambda$CDM model is found in $1\sigma$ regions.     

\begin{widetext}
\begingroup
\squeezetable
\begin{center}
\begin{table}[tbh]
\begin{tabular}{cc|cc|cc|cc}
\hline\hline 
MP & Priors & $\Lambda$wDM (no WiggleZ) & Best fit & $\Lambda$wDM (WiggleZ) & Best fit & $\Lambda$CDM (WiggleZ) & Best fit \\ \hline
$\Omega_b h^2$ & $[0.005,0.1]$ & $0.0220_{-0.000300-0.000579-0.000757}^{+0.000302+0.000600+0.000787}$ & $0.0220$ & $0.0220_{-0.000288-0.000557-0.000738}^{+0.000290+0.000576+0.000762}$ & $0.0222$ & $0.0222_{-0.000242-0.000474-0.000608}^{+0.000242+0.000481+0.000633}$ & $0.0221$\\
$\Omega_c h^2$ & $[0.01,0.99]$ & $0.117_{-0.00170-0.00336-0.00442}^{+0.00169+0.00341+0.00448}$ & $0.117$ & $0.117_{-0.00160-0.00313-0.00408}^{+0.00161+0.00323+0.00428}$ & $0.119$ & $0.118_{-0.00150-0.00291-0.00379}^{+0.00151+0.00295+0.00393}$ & $0.118$\\
$100\theta_{MC}$ & $[0.5,10]$ & $1.0415_{-0.000574-0.00112-0.00145}^{+0.000566+0.00112+0.00147}$ & $1.0414$ & $1.0415_{-0.000572-0.00113-0.00146}^{+0.000571+0.00111+0.00144}$ & $1.0414$ & $1.0416_{-0.000556-0.00108-0.00141}^{+0.000555+0.00109+0.00143}$ & $1.0415$ \\
$\tau$ & $[0.01,0.8]$ & $0.0895_{-0.0143-0.0244-0.0312}^{+0.0122+0.0271+0.0361}$ & $0.0969$ & $0.0893_{-0.0139-0.0243-0.0318}^{+0.0126+0.0263+0.0352}$ & $0.0860$ & $0.0920_{-0.0137-0.0242-0.0314}^{+0.0123+0.0264+0.0350}$ & $0.0839$ \\
$w_{dm}$ & $[-0.2,0.2]$ & $0.00102_{-0.000813-0.00161-0.00211}^{+0.000814+0.00160+0.00208}$ & $0.000858$ & $0.000707_{-0.000747-0.00149-0.00196}^{+0.000746+0.00146+0.00193}$ & $0.000491$ & - & -\\
$n_s$ & $[0.5,1.5]$ & $0.962_{-0.00628-0.0123-0.0165}^{+0.00625+0.0124+0.0163}$ & $0.964$ & $0.963_{-0.00606-0.0119-0.0158}^{+0.00615+0.0121+0.0161}$ & $0.963$ & $0.965_{-0.00550-0.0109-0.0142}^{+0.00551+0.0108+0.0140}$ & $0.965$\\
${\rm{ln}}(10^{10} A_s)$ & $[2.4,4]$  & $3.0902_{-0.0270-0.0467-0.0616}^{+0.0238+0.0513+0.0686}$ & $3.107$ & $3.0887_{-0.0265-0.0467-0.0621}^{+0.0240+0.0497+0.0666}$ & $3.0869$ & $3.0897_{-0.0269-0.0478-0.0622}^{+0.0243+0.0514+0.0679}$ & $3.0713$\\
\hline
$\Omega_\Lambda$ & - & $0.709_{-0.0118-0.0246-0.0331}^{+0.0120+0.0229+0.0297}$ & $0.707$ & $0.704_{-0.0114-0.0235-0.0321}^{+0.0115+0.0220+0.0285}$ & $0.695$ & $0.697_{-0.00872-0.0176-0.0236}^{+0.00948+0.0168+0.0216}$ & $0.698$ \\ 
$\Omega_m$ & - & $0.291_{-0.0120-0.0229-0.0297}^{+0.0118+0.0246+0.0331}$ & $0.293$ & $0.296_{-0.0115-0.0220-0.0285}^{+0.0114+0.0235+0.0321}$ & $0.305$ & $0.303_{-0.00948-0.0168-0.0216}^{+0.00872+0.0176+0.0236}$ & $0.302$ \\
$\sigma_8$ & - & $0.851_{-0.0254-0.0490-0.0644}^{+0.0252+0.0514+0.0682}$ & $0.854$ & $0.843_{-0.0228-0.0444-0.0573}^{+0.0227+0.0461+0.0623}$ & $0.840$ & $0.824_{-0.0121-0.0215-0.0281}^{+0.0110+0.0227+0.0304}$ & $0.816$ \\
$z_{re}$ & - & $11.0229_{-1.0991-2.161-2.836}^{+1.109+2.212+2.898}$ & $11.684$ & $11.00480_{-1.0973-2.146-2.907}^{+1.101+2.154+2.836}$ & $10.738$ & $11.192_{-1.0643-2.131-2.840}^{+1.0786+2.156+2.805}$ & $10.536$\\
$H_0$ & -  & $69.304_{-1.0299-2.0829-2.729}^{+1.0553+2.0665+2.730}$ & $69.0552$ & $68.831_{-0.977-1.924-2.583}^{+0.983+1.952+2.574}$ & $68.186$ & $68.149_{-0.684-1.324-1.735}^{+0.686+1.332+1.728}$ & $68.127$\\
 ${\rm{Age}}/{\rm{Gyr}}$ & - & $13.710_{-0.0657-0.127-0.167}^{+0.0653+0.130+0.170}$ & $13.727$ & $13.738_{-0.0618-0.122-0.157}^{+0.0612+0.123+0.162}$ & $13.756$ & $13.786_{-0.0359-0.0696-0.0890}^{+0.0356+0.0688+0.0905}$ & $13.800$\\
\hline\hline
\end{tabular}
\caption{The mean values with $1,2,3\sigma$ errors and the best fit values of model parameters for general relativity and modified gravity theory, where SNLS3, BAO, {\it Planck}+WMAP9 with and without WiggleZ measurements of matter power spectrum are used.}\label{tab:results}
\end{table}
\end{center}
\endgroup
\end{widetext}

%\begin{widetext}
%\end{widetext}

%\begin{widetext}
%\end{widetext}

%\begin{widetext}
%\end{widetext}

\begin{center}
\begin{figure}[htb]
\includegraphics[width=10cm]{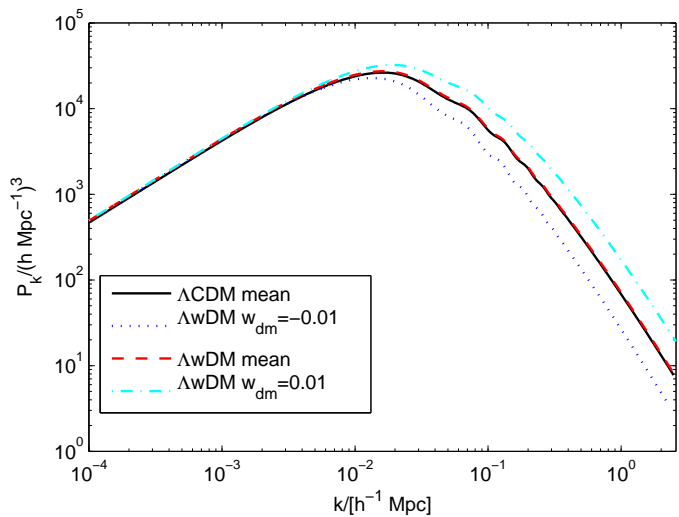}
\caption{The matter power spectra at redshift $z=0$ for different values of the EoS of dark matter $w_{dm}$ where the other relevant model parameters are fixed to their mean values as shown in Table \ref{tab:results}.}\label{fig:mp}
\end{figure}
\end{center}

To show the effects of $w_{dm}$ on the matter power spectra, we fix the other relevant model parameters to their mean values as given in Table \ref{tab:results} and vary the values of $w_{dm}$ from the negative to the positive. Their effects on the matter power spectrum due to different values of $w_{dm}$ are shown in Fig. \ref{fig:mp}, where the redshift is fixed to $z=0$. The constrained result showed the values of $w_{dm}$ is very close to the cold dark matter $w_{dm}=0$; then the linear matter power spectra of them will not be too different as expected. As shown in Fig. \ref{fig:mp}, larger negative values of $w_{dm}$ move the matter and radiation equality to later times and oscillate the matter power spectra (blue, dotted line). And large positive values of $w_{dm}$ move the matter and radiation equality to earlier times and increase the matter power spectrum. Therefore it can be easily understood that hot dark matter is ruled out due to the significant difference on the observations of the large scale structure of our Universe.     
%\begin{widetext}
%\end{widetext}

\section{Conclusion} \label{sec:con}

In this paper, we have investigated the constraints to the equation of state of dark matter by using the currently available cosmic observational data sets, which include the CMB of the first 15.5 months from {\it Planck}, SNLS3, SDSS BAO and WiggleZ measurements of power spectrum. The previous results were updated. We have found that the latest data provide the constraints $w_{dm}=0.000707_{-0.000747}^{+0.000746}$ at $95\%$ C.L.. This result is compatible with the previous results, but a relative tighter constraint was obtained due to the high quality of the currently available data points. The difference of the minimum $\chi^2$ between the $\Lambda$CDM and $\Lambda$wDM models is $\Delta\chi^2_{min}=0.446$ for one extra model parameter $w_{dm}$. Although the currently available cosmic observations favor the $\Lambda$wDM mildly, no significant deviation from the $\Lambda$CDM model is found in the $1\sigma$ region.

\acknowledgements{The authors thank an anonymous referee for helpful improvement of this paper. L. Xu's work is supported in part by NSFC under the Grants No. 11275035 and "the Fundamental Research Funds for the Central Universities" under the Grants No. DUT13LK01.}

\end{document}